\documentclass[rnote]{aa}
\usepackage{graphicx}
\usepackage{amsfonts}
\usepackage{amsmath}
\usepackage{color}
\usepackage{amsbsy}
\usepackage{mathpazo,xfrac}
\usepackage{wasysym}
\usepackage{url,hyperref}
\newlength{\hfwidth}
\newlength{\hfwidthsingle}
\newcommand{\pderiv}[2]{\frac{\partial{#1}}{\partial{#2}}}

\newcommand{\Lu}{\mathrm{Lu}}

\newcommand{\ttimes}[1]{10^{#1}}
\newcommand{\xtimes}[2]{#1\times{10^{#2}}}

\renewcommand{\v}[1]{{\boldsymbol{#1}}} 

\newcommand{\advec}{\left(\v{u}\cdot\del\right)}
\newcommand{\del}{\v{\nabla}}

\newcommand{\Div}{\del\cdot}
\newcommand{\curl}{\del\times}

\newcommand{\va}{v_{_{\rm A}}}

\newcommand{\Eq}[1]{Eq.~(\ref{#1})}

\newcommand{\eq}[1]{\Eq{#1}}

\newcommand{\Fig}[1]{Fig.~\ref{#1}}
\newcommand{\fig}[1]{\Fig{#1}}

\def\blue#1{\textcolor{blue}{#1}}

\begin{document}

\title{Rossby wave instability does not require sharp resistivity gradients}

\author{W. Lyra\inst{1,2,3}, N.J. Turner\inst{1}, \& C.P. McNally\inst{4,5}}
\offprints{wlyra@jpl.nasa.gov}
\institute{Jet Propulsion Laboratory, California Institute of Technology, 4800 Oak Grove Drive, Pasadena, CA, 91109, USA
\and
Department of Geology and Planetary Sciences, California Institute of
Technology, 1200 E. California Ave, Pasadena CA, 91125, USA
\and
Carl Sagan Fellow
\and
Niels Bohr International Academy, Niels Bohr Institute, Blegdamsvej 17, DK-2100, Copenhagen, Denmark
\and
Marie Curie Fellow
}
\date{Received ; Accepted}

\authorrunning{Lyra et al.}
\titlerunning{Wide transition RWI}

\abstract
{ Rossby wave instability (RWI) at dead zone boundaries may 
  play an important role in planet formation. Viscous hydrodynamics 
  results suggest RWI is excited only when the viscosity changes over a
  radial distance less than two density scale heights.  However in the
  disks around Solar-mass T Tauri stars, it is not viscosity but
  magnetic forces that provide the accretion stress beyond about 10 AU,
  where surface densities are low enough so stellar X-rays and
  interstellar cosmic rays can penetrate.} 
{We explore the conditions for RWI in the smooth transition with
increasing distance, from resistive and magnetically-dead to 
conducting and magnetically-active.}
{We perform 3D unstratified MHD simulations with the {\sc Pencil Code}, using static resistivity profiles.}
{We find that in MHD, contrary to viscous models, the RWI is triggered
  even with a gradual change in resistivity extending from 10 to 40\,AU
  (i.e., spanning 15 scale heights for aspect ratio 0.1). This is
  because magneto-rotational turbulence sets in abruptly when the
resistivity reaches a threshold level.  At higher resistivities the
longest unstable wavelength is quenched, resulting in a sharp decline
of the Maxwell stress towards the star. The sharp gradient in
the magnetic forces leads to a localized density bump,
that is in turn Rossby wave unstable.}
{Even weak gradients in the resistivity can lead to sharp transitions in
  the Maxwell stress. As a result the RWI is more
    easily activated in the outer disk than previously
  thought. Rossby vortices at the outer dead zone boundary 
  thus could underlie the dust asymmetries seen in the 
  outer reaches of transition disks.}
\maketitle

\section{Introduction}
\label{sect:introduction}

The RWI (Lovelace \& Hohfeld 1978, Toomre 1981,
Papaloizou \& Pringle 1984, 1985) is a non-axisymmetric hydrodynamic shear instability
associated with axisymmetric extrema of vortensity (Lovelace et
al. 1999, Li et al. 2000). The localized extremum behaves as a potential well that, if deep enough, can trap
modes in its corotational resonance. The trapped modes are linearly
unstable and eventually saturate into large-scale vortices (Hawley
1987, Li et al. 2001, Tagger 2001). Essentially, the RWI is the 
form the Kelvin-Helmholtz instability takes in differentially 
rotating disks, with the upshot being the conversion of
the surplus shear into vorticity. 

The RWI was originally proposed in the context of galaxies (Lovelace
\& Hohfeld 1978), and later, in accretion disk tori (Papaloizou \&
Pringle 1984,1985). First sought as the elusive source of accretion in
disks, the interest in it decreased after the discovery of the 
magnetorotational instability (MRI, Balbus \& Hawley 1991). It has
been reintroduced in the protoplanetary disk landscape by Varni\`ere \&
Tagger (2006) who suggested that the RWI would naturally develop in the
transition between poorly ionized zones that are ``dead'' to the MRI,
and the magnetized zones, ``active'' to the MRI. This is because the
transition constitues a gradient in turbulent viscosity, that in turn
would lead to RWI-unstable density bumps at the dead zone boundaries. 

Subsequently, Lyra et al. (2008b, 2009ab) presented proof-of-concept
models showing that the particle concentration in these vortices
is strong enough to assemble bound clumps of solids, in the Mars-mass
range. Although those models were idealized (2D, no collisional
fragmentation of the particles),
they showed for the first time that planet formation was feasible in
physically-motivated vortices. This scenario has since been shown to hold as increasing
sophistication is sought. The RWI was shown to exist in three
dimensions with vertical stratification in
barotropic (Meheut et al. 2010, 2012ab), polytropic (Lin
2012a), non-barotropic (Lin 2013), and self-gravitating disks (Lin
2012b). Vortices have also been studied in local models, that can more
easily afford high resolution, showing that the ``elliptic
instability'' that destroys an isolated vortex (Kerswell 2002, Lesur \&
Papaloizou 2009) is well balanced by a vorticity source, leading to
vortex survival  (Lesur \& Papaloizou 2010, Lyra \& Klahr 2011).

A critical test was to replace the alpha viscosity
approximation (Shakura \& Sunyaev 1973) by magnetohydrodynamics, with a properly modeled active
zone. This was done by Lyra \& Mac Low (2012) who replaced 
the jump in alpha viscosity by a jump in resistivity, letting the MRI naturally evolve in the
active zone. The RWI was excited, leading to a vortex in the dead side
of the transition. This result was confirmed by Faure et al. (2014),
with a dynamically-varying resistivity profile in a
thermodynamically-evolving disk.

\begin{figure*}
  \begin{center}
    \resizebox{\textwidth}{!}{\includegraphics{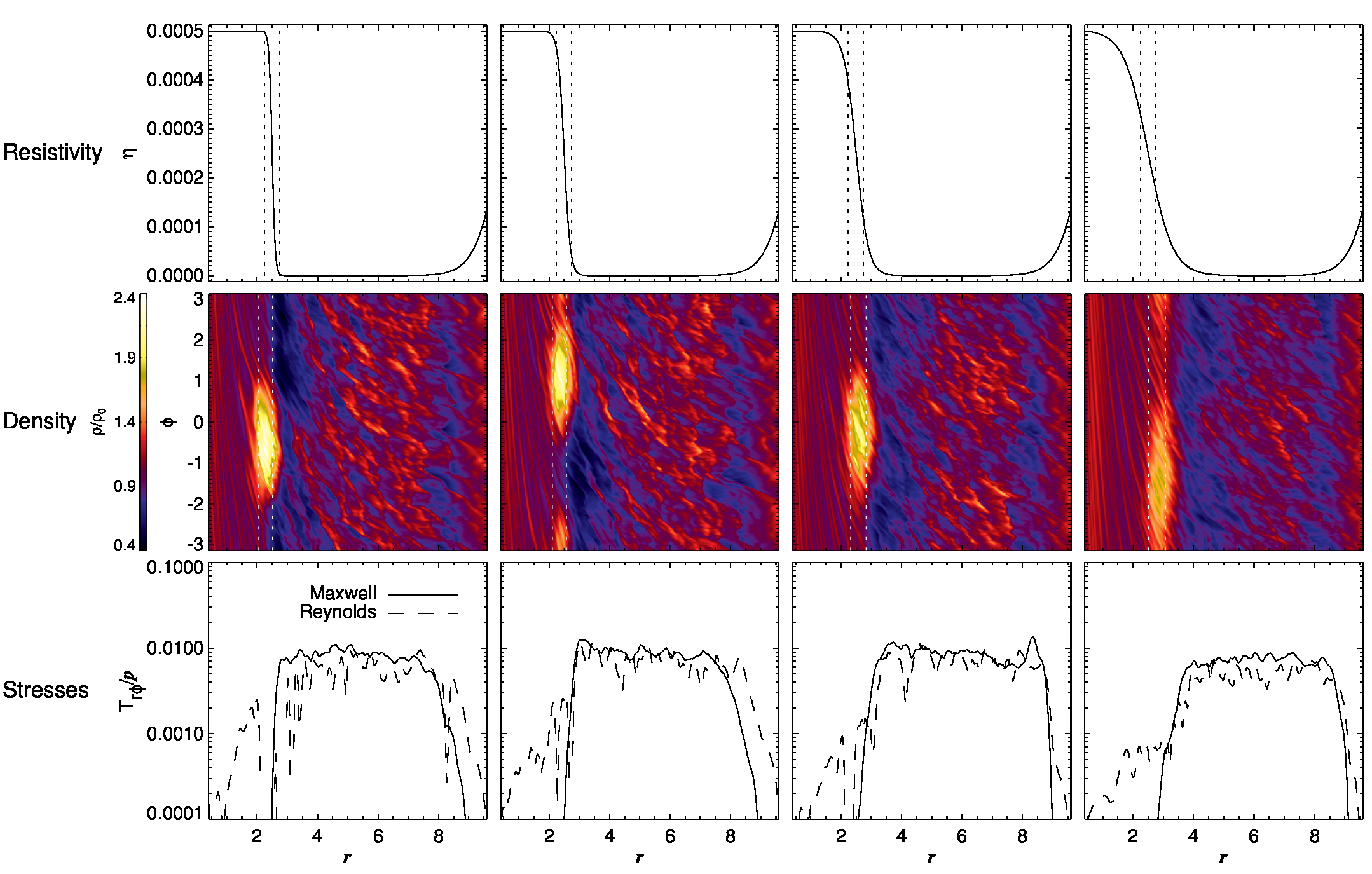}}
\end{center}
\caption[]{Suite of MHD simulations. From left to right, the
  transition width is $h_1$=0.1, 0.2, 0.4 and 0.8. The resulting resistivity
  profiles are shown in the upper panels. The dashed vertical lines
  correspond to $2H$ around the jump center at $r=2.5$. The model with $h_1=0.8$ corresponds to
  a smooth jump from $r\approx 1$ to $r\approx 4$, i.e.,  over
  $\approx$15 scale heights. Although at different times, all models
  develop a sharp density bump that goes unstable to the RWI,
  producing a Rossby vortex (middle panels). 
  The panels, from left to right, correspond to density snapshots at 100, 100, 122,
  and 232$T_0$, where $T_0$ is the orbital period at $r=1$. The
  dashed vertical lines correspond to $2H$ around the density maximum. The
  reason for RWI excitation in these cases is because even though the resistivity jump is smooth, the
  transition in Maxwell stress remains sharp (lower panels). 
  The transition in Reynolds stress is somewhat smoother, but
  turbulent stresses do not blur gradients in the same way Laplacian viscosity
  does, and the density bump remains RWI-unstable.}
\label{fig:results-mhd}
\end{figure*} 

\begin{figure*}
  \begin{center}
    \resizebox{\textwidth}{!}{\includegraphics{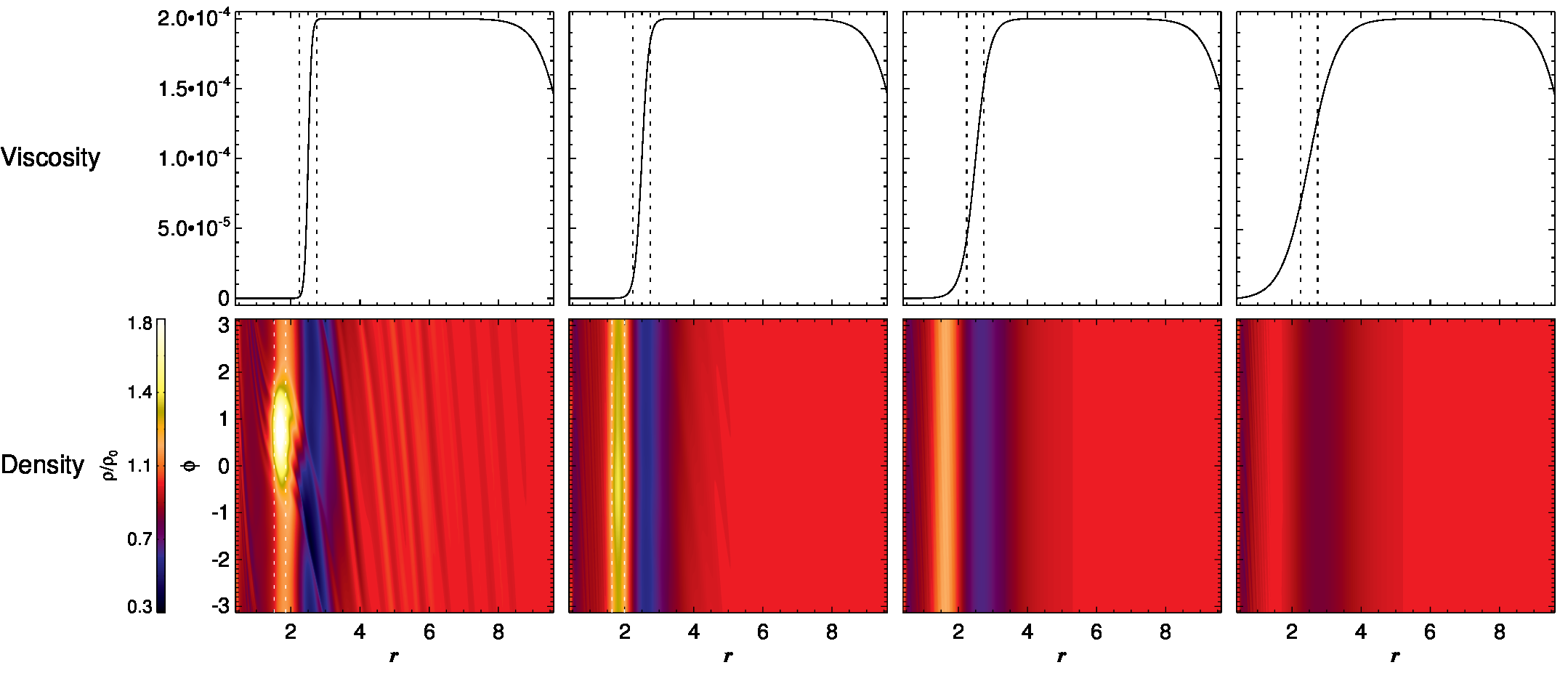}}
\end{center}
\caption[]{Two-dimensional hydrodynamical alpha-disks, with viscosity transitions
  (upper panels) equivalent to the resistivity ones, and $\alpha_\nu\approx 0.02$ in the
  ``active'' zone (the subscript $\nu$ underscores that this
    $\alpha$ corresponds to a  Laplacian viscosity, not to turbulent
    stresses). The lower panels show the density, for snapshots at the same times as in
  \fig{fig:results-mhd}. Only the first ones ($h_1$=0.1 and $h_1$=0.2) develop the RWI.}   
\label{fig:results-visc}
\end{figure*} 

One of the remaining questions toward more realism in this scenario is
how sharp does the transition need to be, and how realistic 
is it to expect such a gradient. Although linear theory predicts that any
extremum of vortensity is unstable to the RWI, one finds in practice
that the extremum has to be sharp enough to trap modes in the
corotational resonance. The required width is problem-dependent, 
but as a rule-of-thumb, Li et al. (2000) find that a 10\%-20\% radial 
variation in density over a length scale comparable to the pressure 
scale height ($H$) is sufficient to trigger it. Lyra et al. (2009a) have shown
that viscous jumps up to $2H$ in width are necessary, with wider jumps
not exciting the RWI up to 500 orbits. This result was subsequently
confirmed by Regaly et al. (2012), up to $\approx$20\,000 orbits. 
Two scale heights is enough in the transition from the inner active to
the outer dead zone, because, in this case, the transition is indeed
sharp. It comes about when the temperature reaches $\approx$900\,K,
enabling collisional ionization of alkali metals (especially
potassium). The model of Lyra \& Mac Low (2012) was
locally-isothermal, using a static sharp resistivity transition
(essentially a Heaviside function), while Faure et al. (2014)
used a thermodynamical model where the resistivity drops to 
zero below the temperature threshold of MRI activation. 

The theme of this paper is the transition in the outer disk. This is
more problematic because there is no sharp threshold. The ionization
increases gradually as X-rays (and perhaps cosmic rays) reach the
midplane. The resistivity thus decreases very smoothly, 
over a very wide range, from $r\approx10$\,AU to $r\approx40$\,AU
(Dzuyrkevich et al. 2013). For aspect ratios $h=H/r=0.1$, this corresponds to about 15 scale heights, and
therefore the RWI is not expected to occur. However, a smooth
  resistivity jump does not necessarily imply an equally smooth
  transition in turbulent stress. In fact, Sano \& Stone (2002) show shearing
  box simulations with net vertical field and resistivity where an
  abrupt transition is seen. In their figure 9 they plot Maxwell stress
  versus Elsasser number, which is a magnetic Reynolds number 
  defined as 

\begin{equation}
  \varLambda \equiv \va^2/\eta\varOmega.
\end{equation} 

\noindent In this definition, 

\begin{equation}
  \va\equiv B/\sqrt{\mu_0\rho}
\end{equation} is the Alfv\'en
  speed, with $\v{B}$ the magnetic field, $\rho$ the density and $\mu_0$ the
  magnetic permitivity of vacuum; $\eta$ is the resistivity and $\varOmega$
  the Keplerian angular frequency. The figure shows that the Maxwell stress is constant for 
  $\varLambda \geq 1$, yet it drops by two orders of magnitude for $\varLambda =
  0.1$. In this work we explore the connection of this result to
  global disk calculations and, in particular, its implications for the RWI.

\section{Model}
\label{sect:model}

\begin{figure*}
  \begin{center}
    \resizebox{\textwidth}{!}{\includegraphics{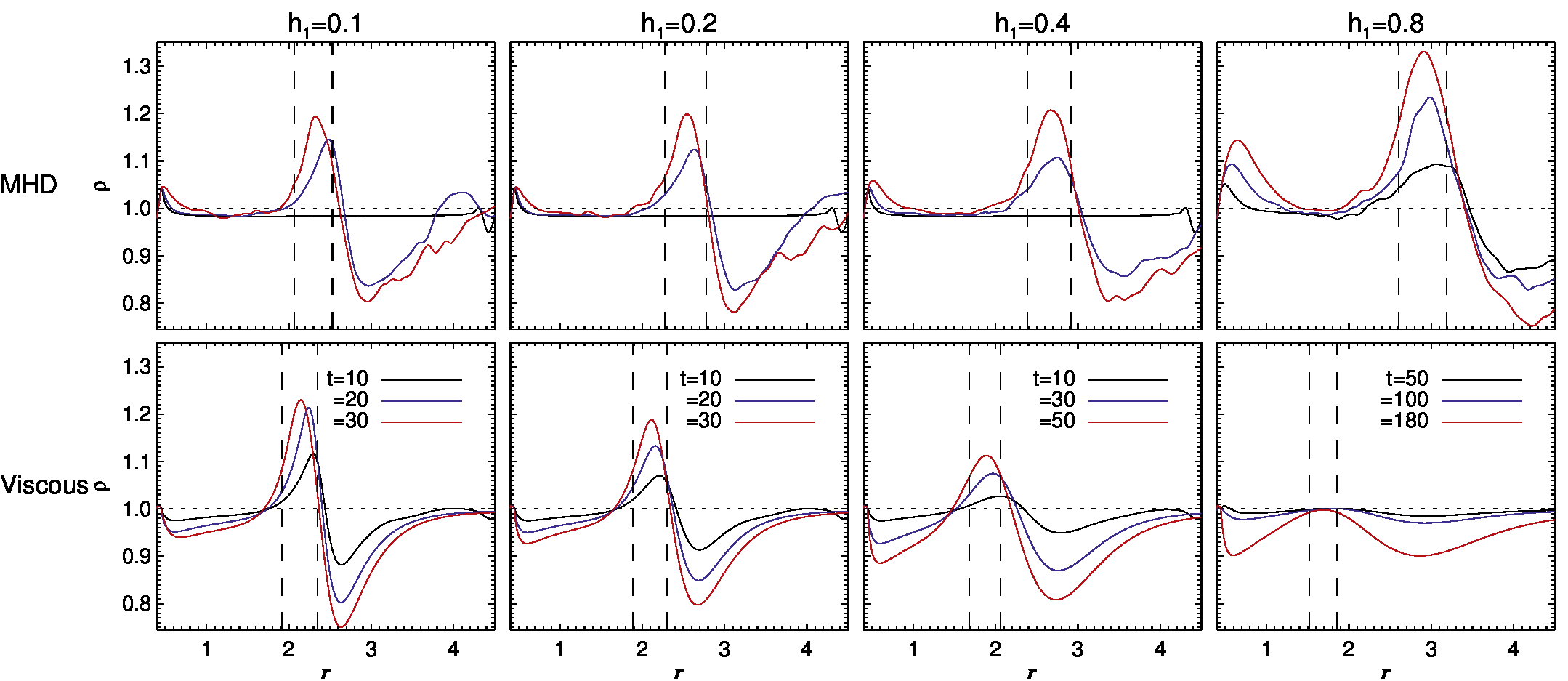}}
\end{center}
\caption[]{Density bumps at selected snapshots, for the MHD
  simulations (upper panels) and alpha-disk equivalents (lower
  panels). The dashed lines bracket $2H$ around the density maximum of
  the last snapshot shown (red line).  The RWI ``rule-of-thumb'' of 10-20\% variation in density
  over a width comparable to the scale height is satisfied for the MHD
  turbulent runs, even for the wide transition model of $h_1=0.8$. In the
  equivalent viscous laminar model the density bumps diffuse over a
  much larger width, and the RWI is not excited.} 
\label{fig:mhdvsvisc}
\end{figure*} 

We perform three-dimensional MHD simulations in the cylindrical 
approximation, i.e., neglecting the disk vertical stratification 
and switching off gravity in that direction. The equations solved are 

\begin{eqnarray}
  \pderiv{\rho}{t}  &=& -\advec\rho -\rho{\Div\v{u}}, \label{eq:continuity}\\
  \pderiv{\v{u}}{t} &=& -\advec\v{u} -\frac{1}{\rho}\del{p} - \del\varPhi + \frac{\v{J}\times\v{B}}{\rho}, \label{eq:navier-stokes}\\
  \pderiv{\v{A}}{t} &=& \v{u}\times\v{B} -\eta\mu_0\v{J} \label{eq:induction}\\
  p&=&\rho c_s^2\label{eq:eos}.
\end{eqnarray}

\noindent where $\v{u}$ the velocity, $\v{A}$ is the 
magnetic potential ($\v{B}=\curl{\v{A}}$), $\v{J}=\mu_0^{-1}\curl{\v{B}}$ is the current density, $p$ is the 
pressure, and $c_s$ is the sound speed. The gravitational potential $\varPhi=-GM_\star/r$ where 
$G$ is the gravitational constant, $M_\star$ is the stellar mass, and 
$r$ is the cylindrical radius. The resistivity is a radial function of 
position. We use smooth step functions 

\begin{equation}
\eta(r) = \eta_0  - \frac{\eta_0}{2}\left[\tanh\left(\frac{r-r_1}{h_1}\right)  -\tanh\left(\frac{r-r_2}{h_2}\right)\right]
\label{eq:eta-jump}
\end{equation}

\noindent in order to mimic the effect of a dead zone. The resistivity passes 
from $\eta_0$ to zero over a width $h_1$ centered at an arbitrarily 
chosen distance $r_1$. The second transition is for buffer 
purposes, raising the resistivity above the MRI-triggering threshold near the outer
boundary of the domain. 

We solve the equations with the {\sc Pencil Code} {\footnote{The code,
including improvements done for the present work, is publicly
available under a GNU open source license and can be downloaded at
http://www.nordita.org/software/pencil-code}} which integrates
the evolution equations with sixth order spatial derivatives, and
a third order Runge-Kutta time integrator. Sixth-order
hyper-dissipation terms are added to \eq{eq:continuity}-\eq{eq:induction},
to provide extra dissipation near the grid scale, explained in Lyra et
al. (2008a). They are needed because the high order scheme of
the Pencil Code has little overall numerical dissipation (McNally et
al. 2012). 

\subsection{Initial Conditions}

We model a disk similar to Lyra \& Mac Low (2012) in cylindrical
coordinates $(r,\phi,z)$. The disk ranges over $r$=[0.4,9.6]$r_0$, 
$\phi$=[-$\pi$,$\pi$], and $z$=$[-0.1,0.1]$. The resolution is 
[$N_r$,$N_\phi$,$N_z$]=[512,512,32]. The radial spacing is
non-uniform, keeping a constant number of
points per radial scale height ($H/\Delta r=16$), where $\Delta
r =\Delta r(r)$ is
the (local) radial resolution. The vertical direction is
unstratified, and the main purpose of its presence is to resolve the
MRI.

We use units such that

\begin{equation}
  GM_\star = r_0 = \varOmega_0 = \rho_0 = \mu_0 = 1. 
  \label{eq:units}
\end{equation}

The density and sound speed are set as radial power-laws 

\begin{equation}
  \rho = \rho_0 \left(\frac{r}{r_0}\right)^{-q_\rho}; \quad \quad c_s^2 = c_{s0}^2 \left(\frac{r}{r_0}\right)^{-q_{_T}} 
\end{equation}

\noindent with $q_\rho=1.5$ and $q_{_T}=1.0$. The reference
sound speed is set at $c_{s0}$=0.1, and Gaussian-distributed noise is added to the
velocities, cell by cell, at rms equal to $\ttimes{-3}$ times the
local sound speed. The initial angular velocity profile is corrected by the thermal pressure gradient

\begin{equation}
  \dot\phi^2 = \varOmega^2 + \frac{1}{r\rho}\frac{\partial{p}}{\partial{r}}
  \label{eq:centrifugal}
\end{equation} \noindent where $\varOmega = \varOmega_0\,(r/r_0)^{-q}$ with $q=1.5$ 
is the Keplerian angular velocity.

The magnetic field is set as a net vertical field, with two MRI wavelengths 
resolved in the vertical range. The constraint $\lambda_{\rm MRI} = 2\pi\va\varOmega^{-1} = L_z/2$ 
translates into a radially varying field 

\begin{equation}
  \va = \frac{L_z\varOmega}{4\pi}\sqrt{\mu_0 \rho} = B_0 \left(\frac{r}{r_0}\right)^{-(q+q_\rho/2)},
\end{equation}

\noindent i.e., a field falling as a 9/4 power-law, with $B_0= \va \approx \xtimes{1}{-2}$ 
in code units. Since the magnetic field is not needed in the
  resistive inner disk, we bring it smoothly to zero inward of $r=2.5$. 
The dimensionless plasma beta parameter $\beta =
2c_s^2/\va^2$ ranges from $\ttimes{3}$ to $\ttimes{4}$ in the active
zone. In this configuration, the MRI grows and saturates 
quickly in 3 local orbits. We use reflective boundaries, with a buffer 
zone of width $H$ at each radial border, that drives the quantities to the initial condition
on a dynamical timescale.

In the presence of resistivity, the behavior of the MRI is
controlled by the magnetic Reynolds number ${\rm Re}_M$$=$$L\va/\eta$ where $L$ is a
relevant length scale. Setting $L=\va/\varOmega$ defines the
Elsasser number $\varLambda$. Because the MRI exists for
arbitrarily large wavelengths (Balbus \& Hawley 1991), a more relevant length scale, that controls 
the excitation of the MRI, is the longest wavelength present in the
vertical domain. For stratified disks, this wavelength is $L=H$. 
In the unstratified case, it is simply $L=L_z$. 
Once this wavelength is in the resistive range, the MRI is
quenched (e.g., Pessah 2010, Lyra \& Klahr 2011). That defines a 
``box'' Lundquist number,

\begin{equation}
  \Lu \equiv \frac{L_z\va}{\eta}.
\end{equation}

We set the reference resistivity $\eta_0$ so that
$\Lu$ is unity at $r_1$, 
thus quenching the MRI inward of this radius. This constraint
translates into $\eta_0 = L_z \va = \xtimes{5}{-4}$. We 
place the resistivity jump starting at $r_1=2.5$, and
  vary the transition width $h_1$. For $h_1=0.8$, the resistivity jump
  goes from $\Lu\approx 0.1$ to 1 from the
inner disk to $r=2.5$, and thence to $\Lu\approx 10$ at
$r\approx 4$. Considering $r_0=10$\,AU, this corresponds to the 
physical dead-to-active transition in the outer disk (Dzyurkevich et
al. 2013). We fix $r_2$=10 and $h_2=0.8$ for
the second transition.

\section{Results}
\label{sect:results}

\begin{figure}
  \begin{center}
    \resizebox{\columnwidth}{!}{\includegraphics{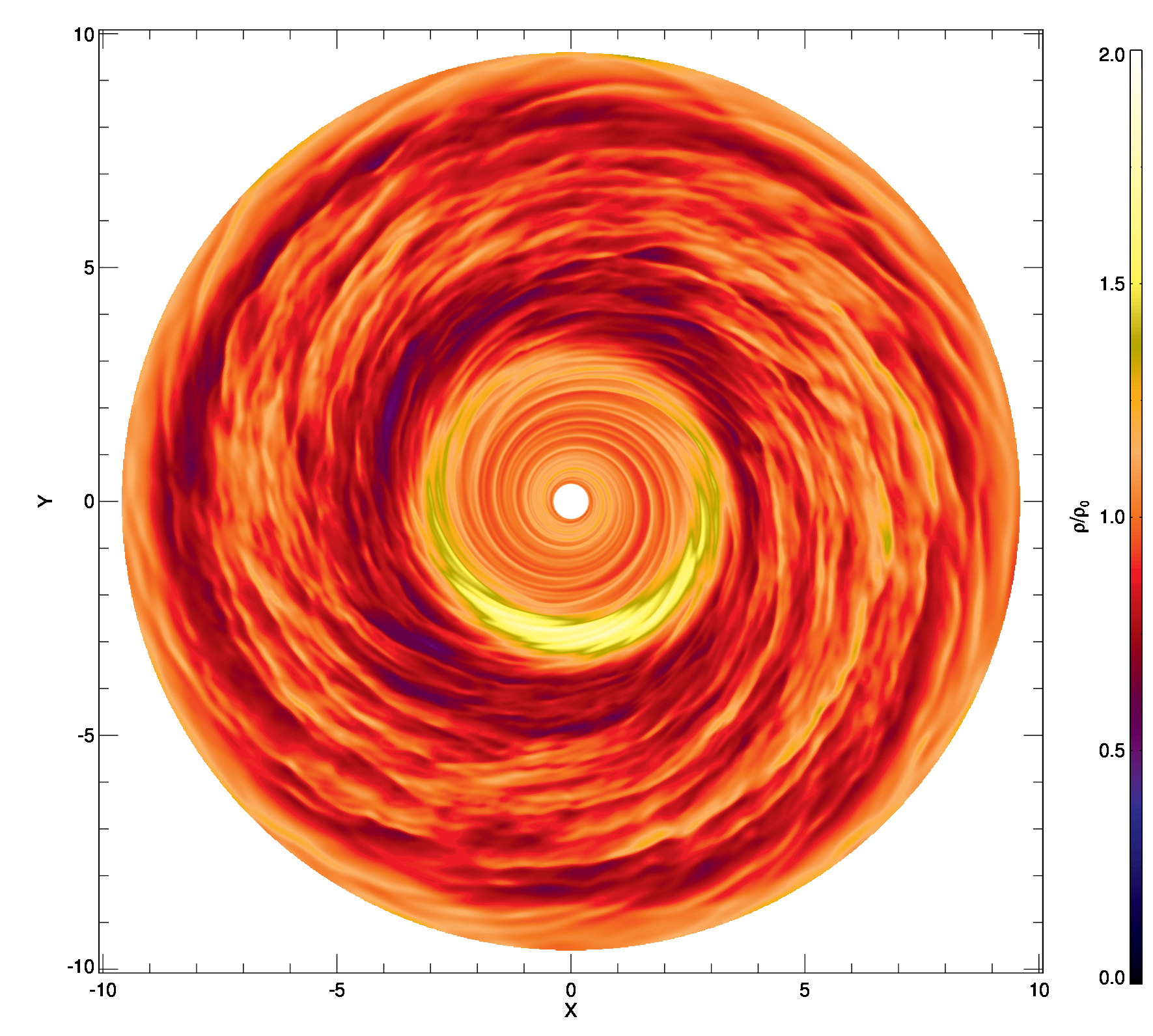}}
\end{center}
\caption[]{Cartesian projection of the MHD model with $h_1=0.8$, with the
  smooth resistivity jump from $r\approx 1$ to $r\approx 4$,
  i.e. roughly 15 scale heights. Although the resistivity jump is
  smooth, the resulting transition in Maxwell stress is
    sharp, triggering the Rossby vortex. Notice also the conspicuous spiral
  pattern propagating into the dead zone. For an animation of the simulation, 
  click \href{https://www.youtube.com/watch?v=XBH5o1q9pZI}{\blue{here}}.}
\label{fig:cartdisk}
\end{figure} 

We show in \fig{fig:results-mhd} the results of the suite of MHD
simulations. The only difference between the models is the width 
of the transition from the (inner) dead zone to the active zone. From left to right, the transition width $h_1$ is 
0.1, 0.2, 0.4, and 0.8. The upper panels show the corresponding
resistivity jumps; the dashed vertical lines mark a width of $2H$ 
around $r=2.5$. For the latter two simulations, the 
transition in resistivity is significantly wider than $2H$. 

The middle panels show the density in the
midplane, normalized by the initial density. The RWI was excited in
all models. The vortex is seen as a non-axisymmetric density
enhancement around (but not exactly at) the resistivity transition
center at $r=2.5$. The white dashed lines correspond to a $2H$
width centered at the density maximum. That the $2H$ lines box the structures
is evidence that they are indeed vortices (whose size is limited by
shocks and therefore approximately $2H$, c.f., Lyra \& Lin 2013).  
The snapshots are at 100, 100, 122, and 232 orbits. 

In the middle panels we show the turbulent alpha values,
i.e., the magnetic and kinetic turbulent stresses normalized by the
local pressure (Shakura \& Sunyaev 1973). We define them, respectively, as 

\begin{equation}
  \alpha_M = \frac{\langle{\va}_r^\prime \
    {\va}_\phi^\prime\rangle}{c_s^2} \qquad \alpha_R = \frac{\langle u_r^\prime \ u_\phi^\prime\rangle}{c_s^2},\\
\end{equation}

\noindent where the angled brackets represent vertical and azimuthal average. The
prime represents a fluctuation from a mean, defined as $\xi^\prime = \xi
- \langle\xi\rangle$, where $\xi$ is an arbitrary quantity. As
  seen in the figure, both stresses level at $\alpha \approx 0.01$. The Maxwell
stress is well confined to the active zone; the Reynolds stress in the
dead zone is roughly an order of magnitude lower than in the active
zone. 

We show in \fig{fig:results-visc} the corresponding 2D 
viscous alpha-disk simulations. The viscosity profile corresponds to that of
the resistivity, but with minima and maxima swapped,

\begin{equation}
  \nu(r) = \frac{\nu_0}{2} \left[\tanh\left( \frac{r-r_1}{h_1}\right) - \tanh\left( \frac{r-r_2}{h_2} \right) \right].
\end{equation}

\noindent The values of $r_1$, $h_1$, $r_2$ and $h_2$ are identical to those
used in the MHD models. The second jump is not needed in the
viscous calculation, but kept for symmetry
reasons. The value of $\nu_0$ is chosen consistently with the value of
$\alpha\approx0.02$ (for the combined kinetic and magnetic stresses) derived
from the MHD calculation. 

The snapshots were taken at the
same times. In agreement with previous alpha-disk works on the RWI,
but in contrast to the MHD runs shown in \fig{fig:results-mhd}, only in the first two simulations, 
where the viscosity jump is sharper than $2H$, was the RWI
triggered. 

\section{Discussion}
\label{sect:discussion}

According to RWI theory, the instability will be triggered if a
quantity $\mathcal{L}$, associated with vortensity, has an extremum
somewhere in the disk. In the context of locally isothermal disks
$\mathcal{L}$ is half the inverse of vortensity

\begin{equation}
  \mathcal{L} = \frac{\rho}{2\omega_z},
\end{equation}

\noindent where $\omega_z$ is the vertical vorticity. Amplitude and sharpness
of the transition also play a role, although a strict criterion has not
been derived. Linear analysis predicts instability if any extremum
exists, but empirically, it is found that critical amplitude and sharpness
exist for the onset of instability. Li et al. (2000) report that as a rule-of-thumb the 
instability is triggered when the density varies by
10\%-20\% over lengths scales comparable to the scale height $H$;
a condition subsequently confirmed in later simulations with different
numerical schemes and different resolutions. 

We show in \fig{fig:mhdvsvisc} that this 
rule-of-thumb is met in the MHD simulations for the
transition widths used. The figure shows, for each width used, the
density bump in the MHD runs and alpha-disk equivalent, in three
different snapshots, labeled in the figure. In the viscous run (lower
panels) the smoothness of the resulting density bump is strongly correlated with
the width of the viscous jump. The same does not happen for the MHD
run, where the density bump is sharp in all cases, and, although the
rate of mass accumulation slows down as the resistivity transition widens, the
smoothness of the bump is only weakly correlated with the width of the
resistivity gradient, remaining sharp in all cases tested. We did not explore further than the transition
width $h_1=0.8$ because, taking $r_0$=10\,AU, that width
corresponds to  a transition over 30\,AU, which is the physical value
expected for the dead-active transition in the outer disk (Dzyurkevich
et al. 2013). The
conclusion is that in physical disks, even the smooth outer resistivity transition
can excite the RWI. 

That even
this smooth transition can lead to excitation of the RWI is in
principle unexpected. How can the density bump be so sharp? The
solution seems to be in the Maxwell stress. Although the resistivity
is smooth, the transition in Maxwell stress remains sharp, a feature that is absent in
the alpha-disk viscous equivalent. The origin of this behavior is a 
property of the MRI. The MRI is excited and maintained for
$\Lu >1$, which always constitutes a sharp transition to turbulence. So, in
essence, it is a property of turbulent flows. As long as the critical
wavelength is not within the resistive range, the MRI will be
excited. This is in agreement with the shearing box results of
  Sano \& Stone (2002). The novelty of this work is to show that this 
non-uniform $\eta-\alpha$ relationship triggers the RWI even for
weak gradients of $\eta$. Maxwell stresses drive the gas inwards (away from
the pressure maximum), placing it at the transition, at
the laminar side. Inviscid, the density bump is slightly blurred by
Reynolds stresses, but does not spread viscously as in the alpha-disk
case, and thus remains sharp. Because the vortex is in the
  resistive side of the transition, it does not get destroyed by
  the magneto-elliptic instability (Mizerski \& Bajer 2009, Lyra \&
  Klahr 2011, Mizerski \& Lyra 2012).

We show in \fig{fig:cartdisk} a Cartesian projection of the
MHD model with $h_1=0.8$. The Rossby vortex is conspicuous as
  a crescent-shaped overdensity at $r\approx 2$. Notice also the spiral
  pattern in the dead zone. Reminiscent of planet-induced spirals, in
  this case the spiral is the result of waves propagating inwards, from the
  turbulence in the active zone.

\section{Conclusions}
\label{sect:conclusions}

The RWI requires a sharp extremum
of potential vorticity. In alpha disks, this bump comes about only at
sharp viscosity transitions, sharper than $2H$. This, although making
the inner active/dead zone transition attractive for the RWI, has hindered the appeal of
the outer dead/active zone transition as a RWI location. 

Sano \& Stone (2002) found that, in shearing boxes, the Maxwell stress
  increases sharply with Elsasser number in the range below unity. Bringing the
  result to global disks, we have found that the required sharpness of the viscosity transition
  for RWI is a feature of alpha disks. We have increased the
width of the resistivity transition without finding a RWI cutoff.  
Resistivity transitions can be as smooth as 30 AU in width
in the outer disk and still excite the RWI. This is because once the
MRI is excited growth rates 
can be affected by resistivity, but the
resulting amplitude at saturation is only weakly affected by it. As a result,
the transition in Maxwell stress remains sharp, driving mass to the dead side of the transition. Once there, without 
Laplacian viscosity to smooth it, the density bump remains sharp,
triggering the RWI as it collects mass. 

Our finding has importance for the interpretation of observations of
dust asymmetries in transitional disks (van der Marel et al. 2013,
Casassus et al. 2013, Isella et al. 2012, 2013, van der Plas et
al. 2014), for which the best explanation is a vortex. A vortex can be brought
about by RWI in gaps carved by planets (de Val Borro 2007, Lyra et
al. 2009b, Lin \& Papaloizou 2011ab, Lin 2012), by convective overstability (Klahr \& Hubbard
2014, Lyra 2014), or by RWI at dead zone boundaries (Varniere \&
Tagger 2006, Lyra et al. 2008b, 2009a). A planetary gap is an exciting possibility (Zhu
\& Stone 2014), but before an undetected planet is invoked, other
alternatives ought to be dealt with. In the case of Oph IRS 48, convective overstability fails to
operate, because the disk is supposedly too radiatively efficient; as
for dead zone boundaries, the transition in resistivity was
thought too smooth to lead to RWI. We have shown that the latter is not a
deterrent. RWI in the outer dead/active transition may be the culprit
for the vortex of Oph IRS 48. 

Notice also the spiral pattern in the dead zone in
\fig{fig:cartdisk}. Similar spiral patterns have been observed in
actual disks (SAO 206462, Muto et al. 2012), and attributed to unseen
planets. In our simulation, these are simply spiral density waves that
propagate inward, launched by the turbulence in the active zone. The
spiral pattern comes about because without turbulent interference,
they propagate coherently. A proper comparison to the
  observations would require the addition of particles in
  order to study how
  they are trapped in these spirals, which we leave for future
  work. At first, we expect the resulting
  particle concentration to be stronger than the trapping 
  in planetary spirals, that are stationary in the reference frame of the
  planet and thus only trap the very well-coupled dust (Lyra et
  al. 2009b). In contrast, these dead zone spirals rotate at the disk's
  velocity and hence should be able to capture more loosely coupled
  particles (Ataiee et al. 2013).

We caution that our models have several limitations.  They are
unstratified MHD calculations with static Ohmic resistivity profiles.
Stratification should have little effect on the linear growth of the
RWI, which is essentially 2-dimensional (Umurhan 2010, Meheut et al.
2012ab, Lin 2012ab, 2013).  However stratification will influence the
outcome by setting a physical scale for the resistivity cutoff: in
unstratified models the results depend on box size, because the
Lundquist number is a function of $L_z$.  In stratified models this
artificial parameter is replaced by the density scale height $H$.
Furthermore the saturated RWI may be severely affected by the vertical
structure.  Lin (2014) finds that Rossby vortices are transient in
disks with a higher-viscosity surface layer (Gammie 1996) if the
density bump spreads out once the vortex forms.  Here the vortex is
long-lived only if the accretion viscosity is low throughout the
column.  However, Lin (2014) also finds that if the viscosity is such
as to maintain the density bump, the vortex is sustained despite the
layered structure.  This latter situation is more similar to our case,
where the turbulent stresses bring material to the dead zone edge,
strengthening the density bump.  Further study of vortex lifetimes in
layered surroundings is warranted, especially treating the ambipolar
and Hall terms, which can greatly alter the vertical structure of the
weakly-ionized annuli in protostellar disks (Wardle 1999, Bai \& Stone
2011, 2013, Wardle \& Salmeron 2012, Mohanty et al. 2013, Kunz \& Lesur 2013, Lesur et al.
2014).  The resistivity's time-dependence has less effect, judging
from the results of Faure et al. (2014).

The layered structure of the magnetic activity potentially has a
special impact at the dead zone edge, where the stratification can
yield a vertically-averaged stress that varies more smoothly with
radius than the midplane profile represented in our unstratified
calculations.  A smoother stress profile would lead to a broader
density bump, weakening our conclusion.  However, dead zone structure
calculations treating the Ohmic and ambipolar terms in prescribed
surface density profiles indicate the transition from dead to active
zone forms a vertical wall up to $\pm H$ (fig. 4a of Dzyurkevich et
al. 2013).  Between $H$ and $2H$, the boundary bends further from the
star, but is set by the ambipolar diffusion, which will weaken as the
density bump builds up.  Over time, mass is thus likely to accumulate
in a narrow range of radii near the position of the dead zone's edge in
the midplane.  The feedback between the evolving density distribution
and the evolving diffusivities deserves further investigation.

Our results are evidence of the practical importance of developing a
detailed picture of the underlying turbulence mechanisms in
protoplanetary disks.  As demonstrated here, such models can have a
critical impact on how observations are understood.

\begin{acknowledgements}
This work was performed in part at the Jet Propulsion
Laboratory, under contract with the California Institute of Technology
funded by the National Aeronautics and Space Administration 
(NASA) through the Sagan Fellowship Program executed by the NASA 
Exoplanet Science Institute. 
The research leading to these results has received funding from the 
People Programme (Marie Curie Actions) of the 
European Union's Seventh Framework Programme 
(FP7/2007-2013) under REA grant agreement 327995.
We acknowledge discussions with Min-Kai Lin and Zhaohuan Zhu.
\end{acknowledgements}


\begin{thebibliography}{}
\bibitem[]{} Ataiee, S., Pinilla, P., Zsom, A., Dullemond, C. P., Dominik, C., \& Ghanbari, J. 2013, A\&A, 553, 3
\bibitem[]{} Bai, X.-N. \& Stone, J. M. 2011, ApJ, 736, 144	
\bibitem[]{} Bai, X.-N. \& Stone, J. M. 2013, ApJ, 769, 76
\bibitem[{{Balbus \& Hawley}(1991)}]{BalbusHawley91} Balbus, S. A. \&  Hawley, J. F. 1991, ApJ, 376, 214
\bibitem[{{Brandenburg \& Dobler}(2002)}]{BrandenburgDobler02} Brandenburg, A. \& Dobler, W. 2002, Comp. Phys. Comm., 147, 471
\bibitem[{{Casassus et al.}(2013)}]{Casassus13} Casassus, S., van der Plas, G., Perez, S., Dent, W.R.F., Fomalont, E., Hagelberg, J., Hales, A., Jord\'an, A., Mawet, D., M\'enard, F., Wootten, A., Wilner, D., Hughes, M., Schreiber, M.R., Girard, J.H., Ercolano, B., Canovas, H., Rom\'an, P., \& Salinas, V. 2013, Nature, 493, 191
\bibitem[{{Dzyurkevich et al.}(2013)}]{Dzyurkevich13} Dzyurkevich. N., Turner, N.J., Henning, Th., \& Kley, W. 2013, ApJ, 765, 114
\bibitem[{{Faure et al.}(2014)}]{Faure2014} Faure, J., Fromang, S., \& Latter, H. 2014, A\&A, 564, 22
\bibitem[]{} Gammie C. F., 1996, ApJ, 457, 355
\bibitem[{{Hawley}(1987)}]{Hawley87} Hawley, J.F. 1987, MNRAS, 225, 677
\bibitem[{{Isella et al.}(2013)}]{Isella13} Isella, A., P\'erez, L.M., Carpenter, J.M., Ricci, L., Andrews, S., \& Rosenfeld, K. 2013, ApJ, 775, 30
\bibitem[{{Isella et al.}(2012)}]{Isella12} Isella, A., P\'erez, L.M., \& Carpenter, J.M. 2012, ApJ, 747, 136
\bibitem[]{} Klahr, H, \& Hubbard, A. 2014, ApJ, 788, 21
\bibitem[]{} Kerswell, R. R. 2002, AnRFM, 34, 83
\bibitem[]{} Kunz, M.W. \& Lesur, G. 2013, MNRAS, 434, 2295
\bibitem[{{Lesur \& Papaloizou}(2009)}]{LesurPapaloizou09} Lesur, G. \& Papaloizou, J.C.B. 2009, A\&A, 498, 1
\bibitem[{{Lesur \& Papaloizou}(2010)}]{LesurPapaloizou10} Lesur, G. \& Papaloizou, J.C.B. 2010, A\&A, 513, 60
\bibitem[]{} Lesur, G., Kunz, M. W., \& Fromang, S. 2014, A\&A, 566, 56
\bibitem[]{}Li, H., Finn, J. M., Lovelace, R. V. E., \& Colgate, S. A. 2000, ApJ, 533, 1023
\bibitem[]{}Li, H., Colgate, S. A., Wendroff, B., Liska, R. 2001, ApJ,
  551, 874
\bibitem[]{} Lin, M.-K. \& Papaloizou, J. C. B 2011a, MNRAS, 415, 1426 
\bibitem[]{} Lin, M.-K. \& Papaloizou, J. C. B 2011b, MNRAS, 415, 1445
\bibitem[{{Lin}(2012a)}]{Lin12} Lin, M.-K. 2012, ApJ, 754, 21
\bibitem[{{Lin}(2012b)}]{Lin12b} Lin, M.-K.\ 2012, MNRAS, 426, 3211
\bibitem[{{Lin}(2013)}]{Lin13} Lin, M.-K. 2013, ApJ, 765, 84
\bibitem[]{} Lin 2014, MNRAS, 437, 575
\bibitem[{{Lovelace \& Hohlfeld}(1978)}]{Lovelace-Hohlfeld78}  Lovelace, R.V.E. \& Hohlfeld, R.G. 1978, ApJ, 221, 51
\bibitem[{{Lovelace et al.}(1999)}]{Lovelace99} Lovelace, R.V.E., Li, H., Colgate, S.A., \& Nelson, A. F. 1999, ApJ, 513, 805
\bibitem[]{} Lyra, W. \& Klahr,  H. 2011, A\&A, 527A, 138
\bibitem[{{Lyra et al.}(2008a)}]{Lyra08a} Lyra, W., Johansen, A., Klahr, H., \& Piskunov, N. 2008a, A\&A, 479, 883
\bibitem[{{Lyra et al.}(2008b)}]{Lyra08b} Lyra, W., Johansen, A., Klahr, H., \& Piskunov, N. 2008b, A\&A, 491, L41
\bibitem[{{Lyra et al.}(2009a)}]{Lyra09a} Lyra, W., Johansen, A., Zsom, A., Klahr, H., Piskunov, N. 2009a, A\&A, 497, 869
\bibitem[{{Lyra et al.}(2009b)}]{Lyra09b} Lyra, W., Johansen, A., Klahr, H., Piskunov, N. 2009b, A\&A, 493, 1125
\bibitem[{{Lyra \& Mac Low}(2012)}]{Lyra-MacLow12} Lyra, W. \& Mac Low, M.-M. 2012, ApJ, 756, 62 
\bibitem[{{Lyra \& Lin}(2013)}]{Lyra-Lin13} Lyra, W. \& Lin, M.-K. 2013, ApJ, 775, 17
\bibitem[]{} Lyra, W. 2014, ApJ, 789, 77
\bibitem[{{McNally et al.}(2012)}]{McNally12} McNally, C. P., Lyra, W., \& Passy, J.-C. 2012, ApJS, 201, 18.
\bibitem[{{M\'eheut et al.}(2010)}]{Meheut10} M\'eheut, H., Casse, F., Varni\`ere, P., \& Tagger M. 2010, A\&A, 516, 31
\bibitem[{{M\'eheut et al.}(2012a)}]{Meheut12a} M\'eheut, H., Cong, Y., \& Lai, D. 2012, MNRAS, 422, 2399
\bibitem[{{M\'eheut et al.}(2012b)}]{Meheut12b} M\'eheut, H., Keppens,  R., Casse, F., \& Benz, W. 2012, A\&A, 542A, 9
\bibitem[]{} Mizerski, K. A., \& Bajer, K. 2009, J. Fluid Mech., 632, 401
\bibitem[]{} Mizerski, K. A., \& Lyra, W. 2012, J. Fluid Mech., 698, 358
\bibitem[]{} Mohanty, S., Ercolano, B., \& Turner, N.J. 2013, ApJ, 764, 65
\bibitem[{{Muto et al.}(2012)}]{Muto12} Muto, T., Grady, C. A., Hashimoto, J., Fukagawa, M., Hornbeck, J. B. et al. 2012, ApJ, 748L, 22
\bibitem[{{Papaloizou \& Pringle}(1984)}]{Papaloizou-Pringle84} Papaloizou, J.C.B. \& Pringle, J.E. 1984, MNRAS, 208, 721
\bibitem[{{Papaloizou \& Pringle}(1985)}]{Papaloizou-Pringle85} Papaloizou, J.C.B. \& Pringle, J.E. 1985, MNRAS, 213, 799
\bibitem[]{} Pessah, M. 2010, ApJ, 716, 1012
\bibitem[{{Reg\'aly et al.}(2012)}]{Regaly12} Reg\'aly, Zs., Juh\'asz,  A., S\'andor, Zs, \& Dullemond, C.P. 2012, MNRAS, 419, 1701
\bibitem[]{} Sano, T. \& Stone, J.M. 2002, ApJ 577, 534
\bibitem[{{Shakura \& Sunyaev}(1973)}]{ShakuraSunyaev73} Shakura, N. I. \& Sunyaev, R. A. 1973, A\&A, 24, 337
\bibitem[{{Tagger}(2001)}]{Tagger01} Tagger, M. 2001, A\&A, 380, 750
\bibitem[{{Toomre}(1981)}]{Toomre81} Toomre, A. 1981, What amplifies the spirals. In {\it The Structure and Evolution of Normal Galaxies}, Proceedings of the Advanced Study Institute, Cambridge, England, Cambridge and New York, Cambridge University Press, 1981, p.111-136.
\bibitem[{{van der Marel et al.}(2013)}]{vanderMarel13} van der Marel,
  N., van Dishoeck, E. F., Bruderer, S., Birnstiel, T., Pinilla, P.,
  Dullemond, C. P., van Kempen, T. A., Schmalzl, M., Brown, J. M., Herczeg, G. J., Matthews, G. S., \& Geers, V. 2013, Science, 340, 1199
\bibitem[{{van der Plas et al.}(2014)}]{vanderPlas14} van der Plas, G., Casassus, S., Ménard, F., Perez, S., Thi, W. F., Pinte, C., \&  Christiaens, V. 2014, ApJ, 792, 25
\bibitem[]{} Umurhan, O.M. 2010, A\&A, 521, 25
\bibitem[{{Varni\`ere \& Tagger}(2006)}]{Varniere-Tagger06} Varni\`ere, P. \& Tagger, M. 2006, A\&A, 446, 13
\bibitem[{{Wardle M.}(1999)}]{} Wardle, M. 1999, MNRAS, 307, 849
\bibitem[{{}()}]{} Wardle M., \& Salmeron R., 2012, MNRAS, 422, 2737
\bibitem[]{} Zhu, Z. \& Stone, J. M. 2014, ApJ, 795, 53 
\end{thebibliography}
\end{document}